\documentclass{article}%
\usepackage{amsmath}
\usepackage{amsfonts}
\usepackage{amssymb}
\usepackage{graphicx}%
\newtheorem{theorem}{Theorem}

\newtheorem{notation}{Notation}

\newenvironment{proof}[1][Proof]{\noindent\textbf{#1.} }{\ \rule{0.5em}{0.5em}}
\begin{document}

\title{The Area Metric Reality Constraint in Classical General Relativity}
\author{Suresh K\ Maran}
\maketitle

\begin{abstract}
A classical foundation for an idea of reality condition in the context of spin
foams (Barrett-Crane models) is developed. I\ extract classical real general
relativity (all signatures) from complex general relativity by imposing the
area metric reality constraint; the area metric is real iff a non-degenerate
metric is real or imaginary. First I\ review the Plebanski theory of complex
general relativity starting from a complex vectorial action. Then I\ modify
the theory by adding a Lagrange multiplier to impose the area metric reality
condition and derive classical real general relativity. I\ investigate two
types of action: Complex and Real. All the non-trivial solutions of the field
equations of the theory with the\ complex action correspond to real general
relativity. Half the non-trivial solutions of the field equations of the
theory with the real action correspond to real general relativity.
Discretization of the area metric reality constraint in the context of
Barrett-Crane theory is discussed. In the context of Barrett-Crane theory the
area metric reality condition is equivalent to the condition that the scalar
products of the bivectors associated to the triangles of a four simplex be
real. The Plebanski formalism for the degenerate case and Palatini formalism
are also briefly discussed by including the area metric reality condition..

\end{abstract}

\section{Introduction}

\subsection{Motivation}

The problem of imposing reality conditions is a non-trivial problem in
canonical quantum gravity \cite{CRBook}. My research indicates that there is
an analogous concept of reality conditions in the context of spin foam models
of gravity \cite{JCBintro}. My goal in this paper is to discuss the classical
foundation of this idea. The quantum application of this idea is dealt with in
Ref:\cite{MSK2}.

Let me briefly discuss ideas from spin foam models which served as the
motivation for this article. Consider the Barrett-Crane models of Lorentzian
general relativity \cite{BCLorentzian}. It is developed using the
Gelfand-Naimarck unitary representation theory of $SL(2,C)$ \cite{IMG}. A
unitary representation of $SL(2,C)$ is labeled by a complex number $\chi
=\frac{n}{2}+i\rho,$ where $\rho$ is a real number and $n$ is an integer. A
Hilbert space $D_{\chi}$ of a unitary representation of the Lorentz group
$SL(2,C)$ is assigned to each triangle of a simplicial manifold. There are two
real Casimirs for $SL(2,C)$. Upto numerical constants the eigenvalues are
$\rho n$ and $-\rho^{2}+\frac{n}{4}^{2}$. The $-\rho^{2}+\frac{n}{4}^{2}$
corresponds to the area spectrum in the Lorentzian Barrett-Crane models. The
Barrett-Crane simplicity constraint requires $\rho n=0$. So we are allowed to
assign only one of $\chi=\rho$ and $\chi=i\frac{n}{2}$ to each triangle.

The two real Casimirs of $SL(2,C)$ can be written together in a complex form
\cite{IMG}:
\[
\hat{C}=\det\left[
\begin{array}
[c]{cc}%
\hat{X}_{3} & \hat{X}_{1}-i\hat{X}_{2}\\
\hat{X}_{1}+i\hat{X}_{2} & -\hat{X}_{3}%
\end{array}
\right]  ,
\]
where $X_{i}=F_{i}+iH_{i}$ $\in$ $sl(2,C)$, the $H_{k}$ correspond to
rotations and the $F_{k}$ correspond to boosts. The eigen-value of the complex
Casimir in $D_{\chi}$ is%
\begin{equation}
\chi^{2}-1=-\rho^{2}+\frac{n}{4}^{2}-1+i\rho n. \label{eq.xarea}%
\end{equation}
The $\rho n$ is precisely the imaginary part of the Casimir. So if $\chi
^{2}-1$ is interpreted as the square of the area of a triangle, then $\rho
n=0$ simply constrains the square of the area to be real.

The reality of the squares of the areas is better understood from the point of
view of the Barrett-Crane model for $SO(4,C)$ general relativity theory
developed in Ref: \cite{MSK2}. The $SO(4,C)\ $Barrett-Crane model can be
constructed using the unitary representation theory of the group $SO(4,C)$
\cite{MSK2}. The unitary representations of $SO(4,C)$ can be constructed using
the relation%
\begin{equation}
SO(4,C)\approx\frac{SL(2,C)\times SL(2,C)}{Z^{2}}. \label{GroupIso}%
\end{equation}
This is the complex analog of
\[
SO(4,R)\approx\frac{SU(2,C)\times SU(2,C)}{Z^{2}}.
\]

So similar to the unitary representation theory of $SO(4,R)$, the unitary
representations of $SO(4,C)$ can be labeled by two `$\chi$'s: $\chi_{L}%
=\frac{n_{L}}{2}+i\rho_{L}$, $\chi_{R}=\frac{n_{R}}{2}+i\rho_{R}$, where each
$\chi$ represents a unitary representation of $SL(2,C)$\footnote{Here the
$n_{L}+n_{R}$ must be an even number. Please see appendix B of \cite{MSK2} for
details.} \cite{IMG}. There are two Casimirs for $SO(4,C)$ which are
essentially the sum and the difference of the Casimirs of the left and the
right handed $SL(2,C)$ parts.

The $SO(4,C)$ Barrett-Crane simplicity constraint sets one of the $SO(4,C)$
Casimir's eigen value $\chi_{L}^{2}-\chi_{R}^{2}$ to be zero, which in turn
sets $\chi_{L}=\pm\chi_{R}$ (=$\chi$ say). Then the other Casimir's eigen
value is
\[
\left(  \chi_{L}^{2}+\chi_{R}^{2}-2\right)  /2=\chi^{2}-1,
\]
which corresponds to the square of the area of a triangle. By setting this
eigenvalue to be real, we deduce the area quantum number to be assigned to a
triangle of a Lorentzian spin foam. So from the point view of the $SO(4,C)$
Barrett-Crane model, the simplicity condition of Lorentzian general relativity
appears to be a reality condition for the squares of the areas.

The Barrett-Crane four simplex amplitude can be formally expressed using a
complete set of orthonormal propagators over a homogenous space of the gauge
group. The $SO(4,C)$ Barrett-Crane model involves the propagators on the
homogenous space $SO(4,C)/SL(2,C)$ which is the complex three sphere $CS^{3}$
\cite{MSK2}. The complex three sphere $CS^{3}$ is defined in $C^{4}$ by%
\[
x^{2}+y^{2}+z^{2}+t^{2}=1,
\]
where $x,y,z,t$ are complex coordinates. The propagators can be considered as
the eigen functions of the square of the area operator with the complex area
eigen values. The homogenous spaces corresponding to real general relativity
theories of all signatures are real subspaces of $CS^{3}$ such that 1) they
possess a complete set of orthonormal propagators\footnote{The propagators are
complete in the sense that there exists a sum over them that yields a delta
function on the homogenous space.} and 2) the propagators correspond to the
real squares of area eigenvalues \cite{MSK2}. Then this naturally suggests
that the spin foams for real general relativity theories for all signatures
are formally related to the $SO(4,C)$ Barrett-Crane model motivated by the
reality of the squares of the areas \footnote{The Barrett-Crane model based on
the propagators on the null-cone \cite{BCLorentzian} is an exception to this.}.

\textbf{Even though the above two paragraphs suggests the reality of the
square of areas as the reality conditions in the context of spin foams the
correct form of the reality conditions will be discussed below.}

\subsection{Content and Organization}

This article aims to develop a classical foundation for the relationship
between real general relativity theories and $SO(4,C)$ general relativity
through a reality constraint which has application to Barrett-Crane theory
\cite{JWBLC1}. The classical continuum analog of the square of area operators
of spin foams is the area metric. In the case of non-degenerate general
relativity, it will be shown in this article that the reality of the area
metric is the necessary and the sufficient condition for real geometry. Since
an area metric can be easily expressed in terms of a bivector $2$-form field,
the area metric reality condition can be naturally combined with the Plebanski
theory \cite{JP1} of general relativity using a Lagrange multiplier.

On a simplicial manifold a bivector two form field can be discretized by
associating bivectors to the triangles. \textbf{In the context of the
Barrett-Crane theory, it will be shown in this article that the necessary and
sufficient condition for the reality of a flat four simplex geometry is
condition that the scalar products of the bivectors associated to the
triangles be real.} This idea in conjunction with the Barrett-Crane constraint
can be used to develop unified treatment of the Barrett-Crane models for the
four dimensional real general relativity theories for all the signatures
(non-degenerate) and $SO(4,C)$ general relativity \cite{MSK2}.

Let me briefly discuss the content and organization of this article. In
section two of this article I\ review the Plebanski formulation \cite{JP1} of
$SO(4,C)$ general relativity starting from vectorial actions. In section three
I discuss the area metric reality constraint. After solving the Plebanski
(simplicity) constraints, I\ show that, the area metric reality constraint
requires the space-time metric to be real or imaginary for the non-denegerate case.

In section four I\ modify the vectorial Plebanski actions by adding a Lagrange
multiplier to impose the reality constraint. For the complex action all the
non-trivial solutions of the field equations correspond to real general
relativity. For the case of real action I show that real general relativity
emerges for non-degenerate metrics for the following cases 1) the metric is
real and the signature type is Riemannian or Kleinien and 2) the metric is
imaginary and Lorentzian.

In section five I discuss the discretization of the area metric reality
constraint on the simplicial manifolds in the context of the Barrett-Crane
theory \cite{JWBLC1} . I\ also discuss various possible discrete actions.

In section six I discuss various further considerations: the area metric
reality constraint for arbitrary metrics, the Plebanski formulation with the
reality constraint for the degenerate case briefly and the Palatini's
formulation with the area metric constraint.

In the appendix I\ have discussed the spinorial expansion of a tensor with the
symmetries of the Riemann curvature tensor.

\section{SO(4,C) General Relativity}

Plebanski's work \cite{JP1} on complex general relativity presents a way of
recasting general relativity in terms of bivector 2-form fields instead of
tetrad fields \cite{APAL} or space-time metrics. It helped to reformulate
general relativity as a topological field theory called the BF theory with a
constraint (for example Reisenberger \cite{MPR1}). Originally Plebanski's work
was formulated using spinors instead of vectors. The vector version of the
work can be used to formulate spin foam models of general relativity
\cite{MPR1}, \cite{LFKK}. Understanding the physics behind this theory
simplifies with the use of spinors. Here I\ would like to review the Plebanski
theory for a $SO(4,C)$ general relativity on a four dimensional real manifold
starting from vectorial actions.

In the cases of Riemannian and $SO(4,C)$ general relativity the Lie algebra
elements are the same as the bivectors. Let me define some notations to be
used in this article.

\begin{notation}
I would like to use the letters $i,j,k,l,m,n$ as $SO(4,C)$ vector indices, the
letters $a,b,c,d,e,f,g,h$ as space-time coordinate indices, the letters
$A,B,C,D,E,F$ as spinorial indices to do spinorial expansion on the coordinate
indices. On arbitrary bivectors $a^{ij}$ and $b^{ij}$, I define $\footnote{The
wedge product in the bivector coordinates plays a critical role in the spin
foam models. This is the reason why the $\wedge$ is used to denote a bivector
product instead of an exterior product.}$
\begin{align*}
a\wedge b  &  =\frac{1}{2}\epsilon_{ijkl}a^{ij}b^{kl}~~\text{and}\\
a\bullet b  &  =\frac{1}{2}\eta_{ik}\eta_{jl}a^{ij}b^{kl}.
\end{align*}

\end{notation}

\subsection{BF $SO(4,C)$ action}

Consider a four dimensional manifold $M$. Let $A$ be a $SO(4,C)$ connection
1-form and $B^{ij}$ a complex bivector valued $2$-form on M. I\ would like to
restrict myself to non-denegerate general relativity in this and the next
section by assuming $b=\frac{1}{4!}\epsilon^{abcd}B_{ab}\wedge B_{cd}\neq0$.
Let $F$ be the curvature 2-form of the connection $A$. I define real and
complex continuum \ $SO(4,C)$ BF\ theory actions as follows,%
\begin{equation}
S_{cBF}(A,B_{ij})=\int_{M}\varepsilon^{abcd}B_{ab}\wedge F_{cd}~~\text{and}
\label{BFcmplx}%
\end{equation}%
\begin{equation}
S_{rBF}(A,B_{ij},\bar{A},\bar{B}_{ij})=\operatorname{Re}\int_{M}%
\varepsilon^{abcd}B_{ab}\wedge F_{cd}. \label{BFreal}%
\end{equation}
The $S_{cBF}$ is considered as a holomorphic functional of it's variables. In
$S_{rBF}$ the variables $A,B_{ij}$ and their complex conjugates are considered
as independent variables. The wedge is defined in the Lie algebra coordinates.
The field equations corresponding to the extrema of these actions are
\begin{align*}
D_{[a}B_{bc]}  &  =0\text{~and}\\
F_{cd}  &  =0.
\end{align*}

$BF$ theories are topological field theories. It is easy to show that the
local variations of solutions of the field equations are gauged out under the
symmetries of the actions \cite{JCBintro}. The spin foam quantization of the
$BF$ theory using the real action has been discussed in Ref:\cite{MSK2}.

\subsection{Actions for $SO(4,C)$ General Relativity}

The Plebanski actions for $SO(4,C)$ general relativity is got by adding a
constraint term to the BF actions. First let me define a complex action
\cite{MPR1},%
\begin{equation}
S_{cGR}(A,B_{ij},\phi)=\int_{M}\left[  \varepsilon^{abcd}B_{ab}\wedge
F_{cd}+\frac{1}{2}b\phi^{abcd}B_{ab}\wedge B_{cd}\right]  d^{4}x,
\label{GRactionComplex}%
\end{equation}
and a real action
\begin{equation}
S_{rGR}(A,B_{ij},\phi,\bar{A},\bar{B}_{ij},\bar{\phi})=\operatorname{Re}%
S_{C}(A,B_{ij},\phi). \label{ComplexAction}%
\end{equation}
The complex action is a holomorphic functional of it's variables. Here $\phi$
is a complex tensor with the symmetries of the Riemann curvature tensor such
that $\phi^{abcd}\epsilon_{abcd}=0$. The $b$ is inserted to ensure the
invariance of the actions under coordinate change.

The field equations corresponding to the extrema of the actions $S_{C}$ and
$S$ are
\begin{subequations}
\label{field equation}%
\begin{align}
D_{[a}B_{bc]}^{ij}  &  =0\text{,}\label{Beq}\\
\frac{1}{2}\varepsilon^{abcd}F_{cd}^{ij}  &  =b\phi^{abcd}B_{cd}^{ij}\text{
and,}\label{Feq}\\
B_{ab}\wedge B_{cd}-b\epsilon_{abcd}  &  =0\text{,} \label{CSeq}%
\end{align}
where $D$ is the covariant derivative defined by the connection $A$. The field
equations for both the actions are the same.

Let me first discuss the content of equation (\ref{CSeq}) called the
simplicity constraint. The $B_{ab}$ can be expressed in spinorial form as
\end{subequations}
\[
B_{ab}^{ij}=B_{AB}^{ij}\epsilon_{\acute{A}\acute{B}}+B_{\acute{A}\acute{B}%
}^{ij}\epsilon_{AB},
\]
where the spinor $B_{AB}$ and $B_{\acute{A}\acute{B}}$ are considered as
independent variables. The tensor
\[
P_{abcd}=B_{ab}\wedge B_{cd}-b\epsilon_{abcd}%
\]
has the symmetries of the Riemann curvature tensor and it's pseudoscalar
component is zero. In appendix A the general ideas related to the spinorial
decomposition of a tensor with the symmetries of the Riemann Curvature tensor
have been summarized. The spinorial decomposition of $P_{abcd}$ is given by%
\begin{align*}
P_{abcd}  &  =B_{(AB}\wedge B_{CD)}\epsilon_{\acute{A}\acute{B}}%
\epsilon_{\acute{C}\acute{D}}+B_{(\acute{A}\acute{B}}\wedge B_{\acute{C}%
\acute{D})}\epsilon_{AB}\epsilon_{CD}+\\
&  \frac{\tilde{b}}{6}\frac{\delta_{c[a}\delta_{b]d}}{2}+B_{AB}\wedge
B_{\acute{A}\acute{B}}(\epsilon_{\acute{A}\acute{B}}\epsilon_{CD}%
+\epsilon_{AB}\epsilon_{\acute{C}\acute{D}}),
\end{align*}
where $\tilde{b}=B_{AB}\wedge B^{AB}+B_{\acute{A}\acute{B}}\wedge B^{\acute
{A}\acute{B}}$. Therefore the spinorial equivalents of the equations
(\ref{CSeq}) are
\begin{subequations}
\label{eq.Pleb}%
\begin{align}
B_{(AB}\wedge B_{CD)}  &  =0,\\
B_{(\acute{A}\acute{B}}\wedge B_{\acute{C}\acute{D})}  &  =0,\\
B_{AB}\wedge B^{AB}+B_{\acute{A}\acute{B}}\wedge B^{\acute{A}\acute{B}}  &
=0~~\text{and}\\
B_{AB}\wedge B_{\acute{A}\acute{B}}  &  =0.
\end{align}
These equations have been analyzed by Plebanski \cite{JP1}. The only
difference between my work (also Reisenberger \cite{MPR1}) and Plebanski's
work is that I\ have spinorially decomposed on the coordinate indices of $B$
instead of the vector indices. But this does not prevent me from adapting
Plebanski's analysis of these equations as the algebra is the same. From
Plebanski's work, we can conclude that the above equations imply $B_{ab}%
^{ij}=$ $\theta_{a}^{[i}\theta_{b}^{j]}$ where $\theta_{a}^{i}$ are a complex
tetrad. Equations (\ref{eq.Pleb}) are not modified by changing the signs of
$B_{AB}$ or/and $B_{\acute{A}\acute{B}}$. These are equivalent to replacing
$B_{ab}$ by $-B_{ab}$ or $\pm\frac{1}{2}\epsilon_{ab}^{cd}B_{cd}$ which
produce three more solution of the equations \cite{LFKK}, \cite{MPR1}.

The four solutions and their physical nature were discussed in the context of
Riemannian general relativity by Reisenberger \cite{MPR1}. It can be shown
that equation (\ref{Beq}) is equivalent to the zero torsion$\ $
condition\footnote{For a proof please see footnote-7 in Ref.\cite{MPR1}.}.
Then $A$ must be the complex Levi-Civita connection of the complex metric
$g_{ab}=\delta_{ij}\theta_{a}^{i}\theta_{b}^{j}$ on $M$. Because of this the
curvature tensor $F_{ab}^{cd}=F_{ab}^{ij}\theta_{i}^{c}\theta_{j}^{c}$
satisfies the Bianchi identities. This makes $F$ to be the $SO(4,C)$ Riemann
Curvature tensor. Using the metric $g_{ab}$ and it's inverse $g^{ab}$ we can
lower and raise coordinate indices. We can define the dualization operation on
an arbitrary antisymmetric tensor $S_{ab}$ as
\end{subequations}
\begin{equation}
\ast S_{ab}=\frac{1}{2}g_{ca}g_{db}\varepsilon^{cdef}S_{ef}, \label{eq.dual}%
\end{equation}
where $\epsilon_{abcd}$ is the undensitized epsilon tensor. It can be verified
that $\ast\ast S_{ab}=gS_{ab}$. To differentiate between the dual operations
on the suffices and the prefixes let me define two new notations:%

\begin{align*}
\underline{S}_{ab}  &  =\ast S_{ab},\\
\overline{S}^{ab}  &  =g^{ac}g^{bd}\ast(g_{ec}g_{fd}S^{ef}).
\end{align*}
Let me assume I have solved the simplicity constraint, and $dB=0$. Substitute
in the action $S$ the solutions $B_{ab}^{ij}=$ $\pm\theta_{a}^{[i}\theta
_{b}^{j]}$ and $A\ $the Levi-Civita connection for a complex metric
$g_{ab}=\theta_{a}\bullet\theta_{b}$. This results in a reduced action which
is a function of the metric only,
\[
S(\theta)=\mp\int d^{4}xbF,
\]
where $F$ is the scalar curvature $F_{ab}^{ab},$and $b^{2}=\det(g_{ab})$. This
is simply the Einstein-Hilbert action for $SO(4,C)$ general relativity.

The solutions $\pm\frac{1}{2}\epsilon_{ab}^{cd}B_{cd}$ do not correspond to
general relativity \cite{LFKK}, \cite{MPR1}. If $B_{ab}^{ij}=$ $\pm\frac{1}%
{2}\epsilon_{ab}^{cd}B_{cd}$, we obtain a new reduced action,%
\[
S(\theta)=\mp\operatorname{Re}\int d^{4}x\epsilon^{abcd}F_{abcd},
\]
which is zero because of the Bianchi identity $\epsilon^{abcd}F_{abcd}=0$. So
there is no other field equation other than the Bianchi identities.

\subsection{Analysis of the field equations}

To extract the content of equation (\ref{CSeq}), let me discuss the spinorial
expansion\footnote{A suitable soldering form and a variable spinorial basis
need to be defined to map between coordinate and spinor space.} of $\phi
_{ab}^{cd}$ and $F_{ab}^{cd}$ $=F_{ab}^{ij}\theta_{i}^{c}\theta_{j}^{d}$.%
\begin{align}
F_{ab}^{cd}  &  =F_{AB}^{CD}\epsilon_{\acute{A}\acute{B}}\epsilon^{\acute
{C}\acute{D}}+F_{\acute{A}\acute{B}}^{\acute{C}\acute{D}}\epsilon_{AB}%
\epsilon^{CD}+\frac{\mathcal{S}}{12}\epsilon_{ab}^{cd}+\frac{\mathcal{F}}%
{12}\delta_{a}^{[c}\delta_{b}^{d]}\\
&  +F_{AB}^{\acute{C}\acute{D}}\epsilon_{\acute{A}\acute{B}}\epsilon
^{CD}+F_{\acute{A}\acute{B}}^{CD}\epsilon_{AB}\epsilon^{\acute{C}\acute{D}%
}~~\text{and}\nonumber
\end{align}%
\begin{align}
\underline{F}_{ab}^{cd}  &  =F_{AB}^{CD}\epsilon_{\acute{A}\acute{B}}%
\epsilon^{\acute{C}\acute{D}}-F_{\acute{A}\acute{B}}^{\acute{C}\acute{D}%
}\epsilon_{AB}\epsilon^{CD}+\frac{\mathcal{F}}{12}\epsilon_{ab}^{cd}%
+\frac{\mathcal{S}}{12}\delta_{a}^{[c}\delta_{b}^{d]}\label{fexp}\\
&  +F_{AB}^{\acute{C}\acute{D}}\epsilon_{\acute{A}\acute{B}}\epsilon
^{CD}-F_{\acute{A}\acute{B}}^{CD}\epsilon_{AB}\epsilon^{\acute{C}\acute{D}%
},\nonumber
\end{align}%
\begin{align}
\underline{\overline{F}}_{ab}^{cd}  &  =F_{AB}^{CD}\epsilon_{\acute{A}%
\acute{B}}\epsilon^{\acute{C}\acute{D}}+F_{\acute{A}\acute{B}}^{\acute
{C}\acute{D}}\epsilon_{AB}\epsilon^{CD}+\frac{\mathcal{S}}{12}\epsilon
_{ab}^{cd}+\frac{\mathcal{F}}{12}\delta_{a}^{[c}\delta_{b}^{d]}\\
&  -F_{AB}^{\acute{C}\acute{D}}\epsilon_{\acute{A}\acute{B}}\epsilon
^{CD}-F_{\acute{A}\acute{B}}^{CD}\epsilon_{AB}\epsilon^{\acute{C}\acute{D}%
},\nonumber
\end{align}
where $\mathcal{F}=F_{ab}^{ab}$ and $\mathcal{S=}\frac{1}{2}\epsilon_{ab}%
^{cd}F_{cd}^{ab}$ . Please notice that in $F_{ab}^{cd},$ the $\mathcal{F}$ and
the $\mathcal{S}$ have exchanged positions due to the dualization. The pseudo
scalar $\mathcal{S}$ is zero since the connection is torsion free.%

\begin{align}
\phi_{ab}^{cd}  &  =\phi_{AB}^{CD}\epsilon_{\acute{A}\acute{B}}\epsilon
^{\acute{C}\acute{D}}+\phi_{\acute{A}\acute{B}}^{\acute{C}\acute{D}}%
\epsilon_{AB}\epsilon^{CD}+\frac{\mathcal{\phi}}{12}\delta_{a}^{[c}\delta
_{b}^{d]}\label{phiexp}\\
&  +\phi_{AB}^{\acute{C}\acute{D}}\epsilon_{\acute{A}\acute{B}}\epsilon
^{CD}+\phi_{\acute{A}\acute{B}}^{CD}\epsilon_{AB}\epsilon^{\acute{C}\acute{D}%
},\nonumber
\end{align}
where $\mathcal{\phi=}\phi_{ab}^{ab}$. The pseudoscalar $\mathcal{\alpha
=}\frac{1}{2}\epsilon_{ab}^{cd}\phi_{cd}^{ab}$ is absent, because it is zero
by definition.

\textbf{Case 1:} $B_{ab}^{ij}=$ $\pm\theta_{a}^{[i}\theta_{b}^{j]}$: In this
case equation (\ref{Feq}) implies
\[
\underline{F}_{ab}^{cd}=b\phi_{ab}^{cd}.
\]
Using the spinor expansions in equations (\ref{fexp}) and (\ref{phiexp}) we
find that the scalar curvature $\mathcal{F}=\mathcal{\alpha=}$ $0$. By
equating the mixed spinor terms and using the exchange symmetry $F_{AB\acute
{C}\acute{D}}=F_{\acute{C}\acute{D}AB}$, we find the trace free Ricci
curvature $F_{\acute{A}\acute{B}}^{CD}$ is zero. Since the scalar curvature
and the trace-free Ricci tensor are the free components of the Einstein
tensor, we have the Einstein's equations satisfied.

\textbf{Case 2:} $B_{ab}^{ij}=\pm\frac{1}{2}\epsilon_{ab}^{cd}\theta_{c}%
^{[i}\theta_{d}^{j]}$: In this case equation (\ref{Feq}) implies
\[
\underline{\overline{F}}_{ab}^{cd}=b\phi_{ab}^{cd}.
\]
Using the spinor expansions we find that there is no restriction on the
curvature tensor $F_{ab}^{cd}$ apart from the Bianchi identities

\section{Reality Constraint for $b\neq0$}

Let the bivector $2$-form field $B_{ab}^{ij}=\pm\theta_{a}^{[i}\theta_{b}%
^{j]}$ and the space-time metric $g_{ab}=\delta_{ij}\theta_{a}^{i}\theta
_{b}^{j}$. Then, the area metric \cite{MPR1} is defined by
\begin{subequations}
\begin{align}
A_{abcd}  &  =B_{ab}\bullet B_{cd}\label{area1}\\
&  =\frac{1}{2}\eta_{ik}\eta_{jl}B_{ab}^{ij}B_{cd}^{kl}\label{area2}\\
&  =g_{a[c}g_{d]b}. \label{area3}%
\end{align}
Consider an infinitesimal triangle with two sides as real coordinate vectors
$X^{a}$ and $Y^{b}$. Its area $A$ can be calculated in terms of the coordinate
bivector $Q^{ab}=\frac{1}{2}X^{[a}Y^{b]}$ as follows%
\end{subequations}
\[
A^{2}=A_{abcd}Q^{ab}Q^{cd}.
\]
In general $A_{abcd}$ defines a metric on coordinate bivector fields:$<\alpha
,\beta>=A_{abcd}\alpha^{ab}\beta^{cd}$ where $\alpha^{ab}$ and $\beta^{cd}$
are arbitrary bivector fields.

Consider a bivector $2$-form field $B_{ab}^{ij}=$ $\pm\theta_{a}^{[i}%
\theta_{b}^{j]}$ on the real manifold $M$ defined in the last section. Let
$\theta_{a}^{i}$ be non-degenerate complex tetrads. Let $g_{ab}=g_{ab}%
^{R}+ig_{ab}^{I}$, where $g_{ab}^{R}$ and $g_{ab}^{I}$ are the real and the
imaginary parts of $g_{ab}=\theta_{a}\bullet\theta_{b}$.

\begin{theorem}
The area metric being real%
\begin{equation}
\operatorname{Im}(A_{abcd})=0, \label{Reality}%
\end{equation}
is the necessary and the sufficient condition for the non-degenerate metric to
be real or imaginary.
\end{theorem}

\begin{proof}
Equation (\ref{Reality}) is equivalent to the following:
\begin{equation}
g_{ac}^{R}g_{db}^{I}=g_{ad}^{R}g_{cb}^{I}. \label{Reality2.2}%
\end{equation}
From equation (\ref{Reality2.2}) the necessary part of our theorem is
trivially satisfied. Let $g,$ $g^{R}$ and $g^{I}$ be the determinants of
$g_{ab},$ $g_{ab}^{R}$ and $g_{ab}^{I}$ respectively. The consequence of
equation (\ref{Reality2.2}) is that $g=g^{R}+g^{I}$. Since $g\neq0$, one of
$g^{R}$ and $g^{I}$ is non-zero. Let me assume $g^{R}\neq$ $0$ and $g_{R}%
^{ac}$ is the inverse of $g_{ab}^{R}$. Let me multiply both the sides of
equation (\ref{Reality2.2}) by $g_{R}^{ac}$ and sum on the repeated indices.
We get $4g_{db}^{I}=g_{db}^{I}$, which implies $g_{db}^{I}$ $=0.$ Similarly
we\ can show that $g^{I}\neq$ $0$ implies $g_{db}^{R}=0$. So we\ have shown
that the metric is either real or imaginary iff the area metric is real.
\end{proof}

\textit{Since an imaginary metric essentially defines a real geometry,
we\ have shown that the area metric being real is the necessary and the
sufficient condition for } \textit{real geometry (non-degenerate) on the real
manifold }$M$. In the last section of this article I discuss this for any
dimensions and rank of the space-time metric.

\section{Extracting Real General Relativity}

To understand the nature of the four volume after imposing the area metric
reality constraint, consider the determinant of both the sides of the equation
$g_{ab}=\theta_{a}\bullet\theta_{b}$,%
\[
g=b^{2},
\]
where $b=\frac{1}{4!}\epsilon^{abcd}B_{ab}\wedge B_{cd}\neq0$. From this
equation we can deduce that $b$ is not sensitive to the fact that the metric
is real or imaginary. But $b$ is imaginary if the metric is Lorentzian
(signature $+++-$ or $---+$) and it is real if the metric is Riemannian or
Kleinien ($++++,----,--++$).

The signature of the metric is directly related to the signature of the area
metric $A_{abcd}=g_{a[c}g_{d]b}$. It can be easily shown that for Riemannian,
Kleinien and Lorentzian geometries the signatures type of $A_{abcd}$ are
$(6,0)$, $(4,2)$ and $(3,3)$ respectively.

Consider the dualizing operator defined in (\ref{eq.dual}) for complex
metrics. Then for real or imaginary metrics it can be verified that
\[
\ast\ast B_{ab}=gB_{ab},
\]
where $g$ $=b^{2}$ is the determinant of the metric.

Consider the Levi-Civita connection
\[
\Gamma_{bc}^{a}=\frac{1}{2}g^{ad}[\partial_{b}g_{cd}+\partial_{c}%
g_{db}-\partial_{d}g_{bc}]
\]
defined in terms of the metric. From the expression for the connection we can
clearly see that it is real even if the metric is imaginary. Similarly the
Riemann curvature tensor
\[
F_{bcd}^{a}=\partial_{\lbrack c}\Gamma_{d]b}^{a}+\Gamma_{b[c}^{e}\Gamma
_{d]e}^{a}%
\]
is real since it is a function of $\Gamma_{bc}^{a}$ only. But $F_{bc}%
^{ad}=g^{de}F_{bce}^{a}$ and the scalar curvature are real or imaginary
depending on the metric.

In background independent quantum general relativity models, areas are
fundamental physical quantities. In fact the area metric contains the full
information about the metric up to a sign\footnote{For example, please see the
proof of theorem 1 of Ref:\cite{TDRKJS}.}. If $B_{ab}^{R}$ and $B_{ab}^{L}$
(vectorial indices suppressed) are the self-dual and the anti-self dual parts
of an arbitrary $B_{ab}^{ij}$, one can calculate the left and right area
metrics as
\[
A_{abcd}^{L}=B_{ab}^{L}\bullet B_{cd}^{L}-\frac{1}{4!}\epsilon^{efgh}%
B_{ef}^{L}\bullet B_{gh}^{L}\epsilon_{abcd}%
\]
and%
\[
A_{abcd}^{R}=B_{ab}^{R}\bullet B_{cd}^{R}+\frac{1}{4!}\epsilon^{efgh}%
B_{ef}^{R}\bullet B_{gh}^{R}\epsilon_{abcd}%
\]
respectively \cite{MPR1}. These metrics are pseudo-scalar component free.
Reisenberger has derived Riemannian general relativity by imposing the
constraint that the left and right area metrics be equal to each other
\cite{MPR1}. This constraint is equivalent to the Plebanski constraint
$B_{ab}\wedge B_{cd}-b\epsilon_{abcd}=0$. I would like to take this one step
further by utilizing the area metric to impose reality constraints on
$SO(4,C)\ $general relativity.

Next, I\ would like to proceed to modify $SO(4,C)$ general relativity actions
defined before to incorporate the area metric reality constraint. The new
actions are defined as follows:
\begin{equation}
S_{c}(A,B,\bar{B},\phi,q)=\int_{M}\varepsilon^{abcd}B_{ab}\wedge F_{cd}%
d^{4}x+C_{S}+C_{R}, \label{Realaction}%
\end{equation}
and
\[
S_{r}(A,B,\bar{A},\bar{B},\phi,\bar{\phi},q)=\operatorname{Re}S(A,B,\bar
{B},\phi,q),
\]
where%
\begin{equation}
C_{S}=\int_{M_{r}}\frac{b}{2}\phi^{abcd}B_{ab}\wedge B_{cd}d^{4}x \label{CS}%
\end{equation}
and%
\begin{equation}
C_{R}=\int_{M}\frac{\left\vert b\right\vert }{2}q^{abcd}\operatorname{Im}%
\left(  B_{ab}\bullet B_{cd}\right)  d^{4}x. \label{CR}%
\end{equation}
The field $\phi^{abcd}$ is the same as in the last section. The field
$q^{abcd}$ is real with the symmetries of the Riemann curvature tensor. The
$C_{R}$ is the Lagrange multiplier term introduced to impose the area metric
reality constraint.

The field equations corresponding to the extrema of the actions under the $A$
and $\phi$ variations are the same as given in section two. They impose the
condition $B_{ab}^{ij}=$ $\pm\theta_{a}^{[i}\theta_{b}^{j]}$ or $\pm\ast
\theta_{a}^{[i}\theta_{b}^{j]}$ and\ $A$ be the Levi-Civita connection for the
complex metric. The field equations corresponding to the extrema of the
actions under the $q^{abcd}$ variations are $\operatorname{Im}(B_{ab}\bullet
B_{cd})=0$. This, as we discussed before, imposes the condition that the
metric $g_{ab}=\theta_{a}\bullet\theta_{b}$ be real or imaginary\footnote{Also
for $B_{ab}^{ij}=\pm\ast\theta_{a}^{[i}\theta_{b}^{j]},$ it can be verified
that the reality constraint implies that the metric $g_{ab}=\theta_{a}%
\bullet\theta_{b}$ be real or imaginary.}.

Let me assume I have solved the simplicity constraint, the reality constraint
and $dB=0$. Substitute the solutions $B_{ab}^{ij}=$ $\pm\theta_{a}^{[i}%
\theta_{b}^{j]}$ and $A\ $the Levi-Civita connection for a real or imaginary
metric $g_{ab}=\theta_{a}\bullet\theta_{b}$ in the action $S$ . This results
in a reduced action which is a function of the tetrad $\theta_{a}^{i}$ only,
\[
S(\theta)=\mp\operatorname{Re}\int d^{4}xbF.
\]
where $F$ is the scalar curvature $F_{ab}^{ab}$. Recall that $F$ is real or
imaginary depending on the metric. This action reduces to Einstein-Hilbert
action if both the metric and space-time density are simultaneously real or
imaginary. If not, it is zero and there is no field equation involving the
curvature $F_{cd}^{ab}$ tensor other than the Bianchi identities.

If $B_{ab}^{ij}=\pm\ast\theta_{a}^{[i}\theta_{b}^{j]}$ , we get a new reduced
action,%
\begin{equation}
S(\theta)=\mp\operatorname{Re}\int d^{4}x\epsilon^{abcd}F_{abcd},
\label{Non-Einstein}%
\end{equation}
which is zero because of the Bianchi identity $\epsilon^{abcd}F_{abcd}=0$. So
there is no other field equation other than the Bianchi identities.

\subsection{Understanding the Field equations}

The field equations corresponding to the extrema of action $S_{r}$ under the
$B$ and $\tilde{B}$ variations about $B_{ab}^{ij}=$ $\pm\theta_{a}^{[i}%
\theta_{b}^{j]}$ are
\begin{subequations}
\label{realeq}%
\begin{align}
\varepsilon^{abcd}F_{cd}^{ij}\varepsilon_{ijkl}  &  =b\phi^{abcd}B_{cd}%
^{ij}\varepsilon_{ijkl}-\frac{i\left\vert b\right\vert b}{2}q^{abcd}%
B_{cd}^{ij}\label{fieldreal1}\\
&  \Longrightarrow\frac{1}{b}\underline{F}_{ab}^{ef}=\phi_{ab}^{~~ef}-\frac
{1}{4}i\bar{q}_{ab}^{~~cd}\left\vert b\right\vert . \label{fieldreal2}%
\end{align}
Here, the star corresponds to dualization on the coordinate variables.

For the action $S_{c},$ only the field equations corresponding to its extrema
under $B$ variations are the same as Eq. (\ref{realeq}). The field equations
corresponding to $\bar{B}$ variations are%
\end{subequations}
\begin{equation}
q^{abcd}B_{cd}^{kl}=0, \label{QzeroEquation}%
\end{equation}
which imply $q^{abcd}$ $=0$ if $b\neq0.$

\subsubsection{The Field Equations of Action $S_{c}$}

Consider the field equation corresponding to the extrema action $S_{c}$ under
the variations of it's variables. Since $q^{abcd}$ $=0$ ($b\neq0$), equation
(\ref{curv.eq}) is the same as equation (\ref{Feq}). So Einstein's equations
are satisfied. Since the metric is essentially real, the field theory of
action $S_{c}$ corresponds to real general relativity.

Please recall that the $b$ is imaginary if the metric is Lorentzian and is
real if the metric is Riemannian or Kleinien. Thus, it is noticed that the
reduced action $S_{c}$ after the reality constraint imposed is real if both
the metric and the space-time density are simultaneously real or imaginary. If
not, the action is imaginary.

\subsubsection{The Field Equations of Action $S_{r}$}

Let me analyze the field equations for action $S_{r}$. Here $q^{abcd}$ need
not be zero.

Let me assume $B_{ab}^{ij}=$ $\pm\theta_{a}^{[i}\theta_{b}^{j]}$, then let me
rewrite equation (\ref{fieldreal2}) below,%

\begin{equation}
\frac{1}{b}\underline{F}_{ab}^{ef}=\phi_{ab}^{~~ef}-\frac{1}{4}i\bar{q}%
_{ab}^{~~cd}\left\vert b\right\vert . \label{curv.eq}%
\end{equation}
There are two different cases now.

\textbf{Case 1:} \textit{The metric and the space-time density }$b$\textit{
are simultaneously real or imaginary}.

Consider the real part of equation (\ref{curv.eq})%
\[
\frac{1}{b}\underline{F}_{ab}^{cd}=\operatorname{Re}\phi_{ab}^{cd}.
\]
This equation is the same as equation (\ref{Feq}) with both the sides being
real. There is no other restriction on $F_{ab}^{cd}$ other than the Bianchi
identities. So Einstein's equations are satisfied. Since $b$ is real, this
case corresponds to Riemannian or Kleinien general relativity.

\textbf{Case 2:} \textit{The metric and the space-time density are not
simultaneously real or imaginary.}

For this case, the imaginary part of equation is (\ref{curv.eq})
\[
\frac{1}{b}\underline{\overline{F}}_{ab}^{cd}=\operatorname{Im}\phi_{ab}%
^{cd}\pm\frac{1}{4}\bar{q}_{ab}^{cd},
\]
with all the terms real. The $q_{ab}^{cd}$ is arbitrary apart from the
constraint imposed by this equation. Therefore we find that there is no
restriction on $F_{ab}^{cd}$ except for the Bianchi identities. So this case
does not correspond to real general relativity.

Let $B_{ab}^{ij}=$ $\pm\ast\theta_{a}^{[i}\theta_{b}^{j]}$. In this case the
field equations corresponding to the extrema of $S_{r}$ under $B_{ab}%
$variations are
\[
\underline{\overline{F}}_{ab}^{cd}=\phi_{ab}^{cd}-i\frac{\left\vert
b\right\vert }{4}\bar{q}_{ab}^{cd}.
\]
This situation is the same as in case (2) of section two, where $F_{ab}^{cd}$
is unrestricted except for the constraints due to Bianchi identities. So this
case does not correspond to general relativity.

\section{Discretization}

\subsection{BF theory}

Consider that a continuum manifold is triangulated with four simplices. The
discrete equivalent of a bivector two-form field is the assignment of a
bivector $B_{b}^{ij}$ to each triangle $b$ of the triangulation. Also the
equivalent of a connection one-form is the assignment of a parallel propagator
$g_{eij}$ to each tetrahedron $e$. Using the bivectors and parallel
propagators assigned to the simplices, the actions for general relativity and
BF theory can be rewritten in a discrete form \cite{FoamDer}. The real
$SO(4,C)$ BF action can be discretized as follows \cite{ooguriBFderv}:%
\begin{equation}
S(B_{b},g_{e})=\operatorname{Re}\sum_{b}B_{b}^{ij}ln{H_{bij}.}
\label{eq.bf.des}%
\end{equation}
The $H_{b}$ is the holonomy associated to the triangle $b$. It can be
quantized to get an spin foam model \cite{MSK2} as done by Ooguri.

\subsection{Barrett--Crane Constraints}

The bivectors $B_{i}$ associated with the ten triangles of a four simplex in a
flat Riemannian space satisfy the following properties called the
Barrett-Crane constraints \cite{JWBLC1}:

\begin{enumerate}
\item The bivector changes sign if the orientation of the triangle is changed.

\item Each bivector is simple.

\item If two triangles share a common edge, then the sum of the bivectors is
also simple.

\item The sum of the bivectors corresponding to the edges of any tetrahedron
is zero. This sum is calculated taking into account the orientations of the
bivectors with respect to the tetrahedron.

\item The six bivectors of a four simplex sharing the same vertex are linearly independent.

\item The volume of a tetrahedron calculated from the bivectors is real and non-zero.
\end{enumerate}

The items two and three can be summarized as follows:
\[
B_{i}\wedge B_{j}=0~\forall i,j,
\]
where $A\wedge B=\varepsilon_{IJKL}A^{IJ}B^{KL}$ and the $i,j$ represents the
triangles of a tetrahedron. If $i=j$, it is referred to as the simplicity
constraint. If $i\neq j$ it is referred as the cross-simplicity constraints.

Barrett and Crane have shown that these constraints are sufficient to restrict
a general set of ten bivectors $E_{b}$ so that they correspond to the
triangles of a geometric four simplex up to translations and rotations in a
four dimensional flat Riemannian space \cite{JWBLC1}.

The Barrett-Crane constraints theory can be easily extended to the $SO(4,C)$
general relativity. In this case the bivectors are complex and so the volume
calculated for the sixth constraint is complex. So we need to relax the
condition of the reality of the volume.

We would like to combine the area metric reality constraint with the
Barrett-Crane Constraints. For this we must find the discrete equivalent of
the area metric reality condition. For this let me next discuss the area
metric reality condition in the context of three simplices and four simplices.
I would like to show that the discretized area metric reality constraint
combined with the Barrett-Constraint constraint requires the complex bivectors
associated to a three or four simplex to describe real flat geometries.

\subsubsection{Three Simplex}

Consider a tetrahedron $t$. Let the numbers $0$ to $3$ denote the vertices of
the tetrahedron. Let me choose the $0$ as the origin of the tetrahedron. Let
$B_{ij}$ be the complex bivector associated with the triangle $0ij$ where $i$
and $j$ denote one of the vertices other than the origin and $i<$ $j$. Let
$B_{0}$ be the complex bivector associated with the triangle $123$. Then
similar to Riemannian general relativity \cite{JWBLC1}, the Barrett-Crane
constraints\footnote{We do not require to use the fifth Barrett-Crane
constraint since we are only considering one tetrahedron of a four simplex.}
for $SO(4,C)$ general relativity imply that
\begin{subequations}
\label{eq.BC}%
\begin{align}
B_{ij}  &  =a_{i}\wedge a_{j},\label{BC1}\\
B_{0}  &  =-B_{12}-B_{23}-B_{34}, \label{BC2}%
\end{align}
where $a_{i}$, $i=1$ to $3$ are linearly independent complex four vectors
associated to the links $0i$ of the three simplex. Let me choose the vectors
$a_{i}$, $i=1$ to $3$ to be the complex vector basis inside the tetrahedron.
Then the complex $3D$ metric inside the tetrahedron is
\end{subequations}
\begin{equation}
g_{ij}=a_{i}\cdot a_{j}, \label{eq.metric}%
\end{equation}
where the dot is the scalar product on the vectors. This describes a flat
complex three dimensional geometry inside the tetrahedron. The area metric is
given by
\[
A_{ijkl}=g_{i[k}g_{l]j}.
\]
The coordinates of the vectors $a_{i}$ are simply%
\begin{align*}
a_{1}  &  =(1,0,0),\\
a_{2}  &  =(0,1,0),\\
a_{3}  &  =(0,0,1).
\end{align*}
Because of this all of the six possible scalar products made out of the
bivectors $B_{ij}$ are simply the elements of the area metric. From the
discussion of the last section the reality of the area metric simply requires
that the metric $g_{ij}$ be real or imaginary. Since $B_{0}$ is also defined
by equation (\ref{BC2}) its inner product with itself and other bivectors are
real. Thus in the context of a three simplex, the discrete equivalent of the
area metric reality constraint is that the all possible scalar products of
bivectors associated with the triangles of a three simplex be real.

\subsubsection{Four Simplex}

In the case of a four simplex $s$ there are six bivectors $B_{ij}$. There are
four $B_{0}$ type bivectors. Let $B_{i}$ denote the bivector associated to the
triangle made by connecting the vertices other than the origin and vertex $i$.
The Barrett-Crane constraints imply equation (\ref{BC1}) with $i,j=$ $1$ to
$4$. There is one equation for each $B_{i}$ similar to equation (\ref{BC2}).
Now the metric $g_{ij}=a_{i}\cdot a_{j}$ describes a complex four dimensional
flat geometry inside the four simplex $s$. Now assuming we are dealing with
non-degenerate geometry, the reality of the geometry requires the reality of
the area metric. Similar to the three dimensional case, the components of the
area metric are all of the possible scalar products made out of the bivectors
$B_{ij}$. The scalar products of the bivectors $B_{i}$ among themselves or
with $B_{ij}$'s are simple real linear combinations of the scalar products
made from $B_{ij}$'s. So one can propose that the discrete equivalent of the
area metric reality constraint is simply the condition that the scalar product
of these bivectors be real. Let me refer to the later condition as the
bivector scalar product reality constraint.

\begin{theorem}
The necessary and sufficient conditions for a four simplex with real
non-degenerate flat geometry are 1) The $SO(4,C)$ Barrett-Crane
constraints\footnote{The $SO(4,C)$ Barrett-Crane constraints differ from the
real Barrett-Crane constraints by the following:
\par
\begin{enumerate}
\item The bivectors are complex, and
\par
\item The condition for the reality of the volume of tetrahedron is not
required.
\end{enumerate}
} and 2) The reality of all possible bivector scalar products.
\end{theorem}

\begin{proof}
The necessary condition can be shown to be true by straight forward
generalization of the arguments given by Barrett and Crane \cite{JWBLC1} and
application of the discussions in the last paragraph. The sufficiency of the
conditions follow from the discussion in the last paragraph.
\end{proof}

The quantization of a four simplex using the $SO(4,C)$ Barrett-Crane
constraints and the bivector scalar product reality constraint has been argued
in Ref.\cite{MSK2}

\subsection{Actions for Simplicial General Relativity}

Here we would like define actions for general relativity which has application
for the Barrett-Crane models \cite{JWBLC1}, \cite{MSK2}.

The discrete BF theory described in equation (\ref{eq.bf.des}) can be further
modified by imposing the $SO(4,C)$ Barrett-Crane constraints on it to get the
$SO(4,C)$ Barrett-Crane model \cite{MSK2}, \cite{FoamDer}. The resulting model
can be considered as a path-integral quantization of the simplicial version of
the action in equation (\ref{Realaction}),
\begin{equation}
S_{GR}(B_{b},g_{e},\phi)=\sum_{b}B_{b}^{ij}ln{H_{bij}}+\frac{1}{2}%
\sum_{b\grave{b}}\phi_{b\grave{b}}B_{b}\wedge B_{\grave{b}},\label{dcrtPleb}%
\end{equation}
where $\phi_{lb\grave{b}}$ are to impose the Barrett-Crane constraints (2) and
(3) on $B_{b}$. There is one $\phi_{b\grave{b}}$ for every pair of triangles
$b\grave{b}$ such that either they are the same or they intersect at a link.

A proposal for an action for real general relativity is a modified form of
equation (\ref{Realaction}) that includes extra Lagrange multipliers to impose
the bivector scalar product conditions\footnote{The square of area reality
conditions state that,
\par
\begin{itemize}
\item the square of the area of the triangle calculated as scalar product of
the associated bivector is real.
\par
\item the square of area calculated as scalar product of sum of the bivectors
associated with two triangle of a tetrahedron is real.
\end{itemize}
\par
Assume the first constraint is imposed on each of any two triangles of a
tetrahedron. Then the second constraint is equivalent to the condition that
the scalar product of the bivectors associated to these triangle is real.}:%
\begin{align}
S_{rGR}(B_{b},g_{e},\phi,q)  &  =\operatorname{Re}\sum_{b}B_{b}^{ij}%
ln{H_{bij}}\label{dcrtreal}\\
&  +\frac{1}{2}\operatorname{Re}\sum_{b\grave{b}}\phi_{b\grave{b}}%
B_{bij}\wedge B_{\grave{b}}^{ij}\nonumber\\
&  +\frac{1}{2}\sum_{b\grave{b}}q_{b\grave{b}}\operatorname{Im}(B_{b}\circ
B_{\grave{b}}),\nonumber
\end{align}
where there is one real $q_{b\grave{b}}$ for every pair of triangles
$b\grave{b}$ such that either they are same or they intersect at a link. {}The
Lagrange multipliers $q_{b\grave{b}}$ helps impose the conditions that

\begin{itemize}
\item the scalar product of a bivector $B_{b}$ with itself is real and

\item the scalar product of a bivectors associated to triangles which
intersect at a link is real.
\end{itemize}

Above we have ignored to impose reality of the scalar products of the
bivectors associated to any two triangles of the same four simplex which
intersect at only at one vertex. This is because these constraints appears not
to be needed for a formal extraction \cite{MSK2} of the Barrett-Crane models
of real general relativity from that of $SO(4,C)$ general relativity. Imposing
these constraints may not be required because of the enormous redundancy in
the bivector scalar product reality constraints defined in the last
section\footnote{Please notice that only about ten independent conditions are
required to reduce a complex four metric to a real four metric.}. This issue
need to be carefully investigated.

An alternative discrete action for general relativity is that of Regge
\cite{ReggeCalc}. In any dimension $n$, given a simplicial geometry, the Regge
action is%

\[
S_{\text{Reg}}=\sum_{b}A_{b}\varepsilon_{b}.
\]
The asymptotic limit of the $SO(4,C)$ Barrett-Crane model recovers
$SO(4,C)\ $Regge Calculus and the bivectors that satisfy the Barrett-Crane
constraints \cite{MSK2}. This is also true for models of real general
relativity theories for various signatures as they are simple restrictions of
$SO(4,C)$ ideas \cite{MSK2}.

Above, the $A_{b}$ are the areas of the triangles expressed as functions of
link lengths of the four simplex. The link lengths are the free variables of
the Regge theory. The $\varepsilon_{b}$ is the deficit around a bone $b$
\cite{ReggeCalc}. This action can be easily generalized to $SO(4,C)$ general
relativity. Similar to the action in equation (\ref{dcrtreal}) the reality
constraints can be combined with the Regge Calculus:%
\begin{equation}
S_{r\text{Reg}}=\sum_{b}A_{b}\varepsilon_{b}.+\frac{1}{2}\sum_{b\grave{b}%
}q_{b\grave{b}}\operatorname{Im}(B_{b}\circ B_{\grave{b}}), \label{eq.regreal}%
\end{equation}
where the $B_{b}$, $A_{b}$ and $\varepsilon_{b}$ can be considered as the
functions of complex vectors associated to the links of the triangulation. The
link vectors can be considered as the free variables of this theory.

In a discrete general relativity theory on the simplicial manifolds we do not
require the continuity of the metric a priori. This means that the flat
geometry associated to each four simplex can be of any signature. This means
that the actions (\ref{dcrtreal}) and (\ref{eq.regreal}) describe a
multi-signature discrete general relativity where the geometry of each simplex
has a different signature \cite{MSK2}.

\section{Further Considerations}

\subsection{Reality Constraint for Arbitrary Metrics}

Here we\ analyze the area metric reality constraint for a metric $g_{ac}$ of
arbitrary rank in arbitrary dimensions, with the area metric defined as
$A_{abcd}=$ $g_{a[c}g_{d]b}$. Let the rank of $g_{ac}$ be $r$.

If the rank $r=1$ then $g_{ab}$ is of form $\lambda_{a}\lambda_{b}$ for some
complex non zero co-vector $\lambda_{a}$. This implies that the area metric is
zero and therefore not an interesting case.

Let me prove the following theorem.

\begin{theorem}
If the rank $r$ of $g_{ac}$ is $\geq2,$then the area metric reality constraint
implies the metric is real or imaginary. If the rank $r$ of $g_{ac}$ is equal
to $1,$ then the area metric reality constraint implies $g_{ac}=\eta\alpha
_{a}\alpha_{b}$ for some complex $\eta\neq0$ and real non-zero co-vector
$\alpha_{a}$.
\end{theorem}

The area metric reality constraint implies%
\begin{equation}
g_{ac}^{R}g_{db}^{I}=g_{ad}^{R}g_{cb}^{I}. \label{eq.reality}%
\end{equation}
Let $g_{AC}$ be a $r$ by $r$ submatrix of $g_{ac}$ with a non zero
determinant, where the capitalised indices are restricted to vary over the
elements of $g_{AC}$ only. Now we have%
\begin{equation}
g_{AC}^{R}g_{DB}^{I}=g_{AD}^{R}g_{CB}^{I}. \label{eq.reality1}%
\end{equation}
From the definition of the determinant and the above equation we have
\[
\det(g_{AC})=\det(g_{AC}^{R})+\det(g_{AC}^{I}).
\]
Since $\det(g_{AC})\neq0$ we have either $\det(g_{AC}^{R})$ or $\det
(g_{AC}^{I})$ not equal to zero. Let me assume $g_{ac}^{R}\neq0$. Then
contracting both the sides of equation (\ref{eq.reality1}) with the inverse of
$g_{AC}^{R}$ we find $g_{DB}^{I}$ is zero. Now from equation (\ref{eq.reality}%
) we have%
\begin{equation}
g_{AC}^{R}g_{dB}^{I}=g_{Ad}^{R}g_{CB}^{I}=0. \label{eq.1}%
\end{equation}
Since the Rank of $g_{AC}^{R}\geq2$ we can always find a $g_{AC}^{R}$ $\neq0$
for some fixed $A$ and $C.$ Using this in equation (\ref{eq.1}) we find
$g_{dB}^{I}$ is zero. Now consider the following:%
\begin{equation}
g_{AC}^{R}g_{db}^{I}=g_{Ad}^{R}g_{Cb}^{I}. \label{eq.2}%
\end{equation}
we can always find a $g_{AC}^{R}\neq0$ for some fixed $A$ and $C.$ Using this
in equation (\ref{eq.2}) we find $g_{db}^{I}=0$. So we have shown that if
$g_{ac}^{R}\neq0$ then $g_{db}^{I}=0$. Similarly if we can show that if
$g_{ac}^{I}\neq0$ then $g_{db}^{R}=0$.

\subsection{The Plebanski Formulation for $b=0$}

The degenerate case corresponding to $b=0$ has been analyzed in the context of
Riemannian general relativity by Reisenberger \cite{MPR1}. In his analysis the
simplicity constraint yields
\[
B_{L}^{I}=T_{J}^{I}B_{R}^{J},
\]
where $B_{L}^{I}$ and $B_{R}^{J}$ are the left handed and the right handed
components of the real bivector valued two-form $B_{ij}$, the integers $I$,
$J$ are the Lie algebra indices and $T_{J}^{I}$ is an $SO(3,R)$ matrix. If the
action is gauge invariant under $SO(4,R),$ in a proper gauge $B_{L}^{I}%
=T_{J}^{I}B_{R}^{J}$ reduces to $B_{L}^{I}=B_{R}^{I}$. Let me denote
$B_{L}^{I}=B_{R}^{I}$ simply by $\Sigma^{I}$. Reisenberger starts from the
Riemannian version of the action in equation (\ref{ComplexAction}) and finally
ends up with the following reduced actions:%
\[
S_{DG}(\Sigma^{I},A_{R},A_{L})=\int_{M_{r}}\delta_{IJ}\Sigma^{I}(F_{R}^{J}\pm
F_{L}^{J}),
\]
where $\Sigma^{I}$ is a $SU(2)$ Lie-algebra valued two form, $A_{R}(A_{L})$ is
a right (left) handed $SU(2)$ connection and $F_{R}$ ($F_{L}$) are their
curvature two forms. This action and the analysis that led to this action as
carried out done in Ref:\cite{MPR1} can be easily generalized to $SO(4,C)$
general relativity by replacing $SU(2)$ with $SL(2,C)$.

Now, in case the of $b=0$ the area metric defined in terms of $B_{ab}^{ij}$ is%
\begin{align*}
A  &  =\frac{1}{2}\eta_{ik}\eta_{jl}B^{ij}\otimes B^{kl}\\
&  =\delta_{IJ}B_{R}^{I}\otimes B_{R}^{J}+\delta_{IJ}B_{L}^{I}\otimes
B_{L}^{J}\\
&  =2\delta_{IJ}\Sigma^{I}\otimes\Sigma^{J}.
\end{align*}
Now the $B$ field is no longer related to a tetrad, which means we do not have
a space-time metric defined. But it can be clearly seen that the area metric
is still defined.

The reduced versions of actions $S_{r}$ and $S_{c}$ for $b=0$ with simplicity
constraint imposed are,%
\[
S_{rDG}(A_{R},A_{L},\Sigma,\bar{\Sigma})=\int_{M_{r}}\varepsilon^{abcd}%
\delta_{IJ}\Sigma_{ab}^{I}(F_{cdR}^{J}\pm F_{cdL}^{J})+\int q^{abcd}%
\operatorname{Im}(\delta_{IJ}\Sigma_{ab}^{I}\Sigma_{cd}^{J})\text{ and}%
\]%
\[
S_{cDG}(\Sigma,A_{R},A_{L},\bar{\Sigma},\bar{A}_{R},\bar{A}_{L}%
)=\operatorname{Re}S_{DG}(\Sigma,A_{R},A_{L},\bar{\Sigma})\text{.}%
\]
The field equations relating to $S_{rDG}$ extrema are
\begin{align*}
D_{R}\Sigma^{I}  &  =D_{L}\Sigma^{I}=0,\\
\frac{1}{2}\epsilon^{abcd}F_{cdR}^{J}  &  =q^{abcd}\Sigma_{cd}^{I},\\
\frac{1}{2}\epsilon^{abcd}F_{cdL}^{J}  &  =-q^{abcd}\Sigma_{cd}^{I}\text{
and}\\
\operatorname{Im}(\delta_{IJ}\Sigma_{ab}^{I}\Sigma_{cd}^{J})  &  =0.
\end{align*}
For $S_{cDG},$ we have additional equations
\[
q^{abcd}\Sigma_{cd}^{I}=0,
\]
which imply
\[
F_{cdR}^{J}=F_{cdL}^{J}=0.
\]

The reality constraint requires $A=2\delta_{IJ}\Sigma^{I}\otimes\Sigma^{J}$ to
be real. Such expression allows for assigning a real square of area values to
the two surfaces of the manifold. The spin foam quantization of the theory of
$S_{rDG}$ without the reality constraint in the case of Riemannian general
relativity has been studied by Perez \cite{AP3}. The spin foam quantization of
the $SO(4,C)$ theory with the reality constraint needs to be studied.

\subsection{Palatini Formalism with the Reality Constraint}

Consider alternative actions of Palatini's form \cite{APAL} which use the
co-tetrads $\theta^{i}$ instead of the bivector $2$-form field as a basic
variable. The Palatini actions with the reality constraint included are
\begin{align*}
S_{cPT}[A,\theta^{i},\bar{\theta}^{i},q^{abcd}]  &  =\int\epsilon_{ijkl}%
\theta^{i}\theta^{j}F^{kl}+q^{abcd}\operatorname{Im}(B_{ab}\bullet
B_{cd})\text{ and}\\
S_{rPT}[\theta^{i},\bar{\theta}^{i},A,\bar{A},q^{abcd}]  &  =\operatorname{Re}%
S[A,\theta^{i},\bar{\theta}^{i},q^{abcd}],
\end{align*}
where $F^{ij}$ is the curvature 2-form corresponding to the $SO(4,C)$
connection $A$ and $B_{ab}=\theta_{a}\wedge\theta_{b}$. The equations of
motion for the theory of $S_{rPT}$ are%
\begin{equation}
D(\theta^{k}\theta^{l})=0, \label{three}%
\end{equation}%
\begin{equation}
\epsilon^{abcd}\epsilon_{jkl}^{i}\theta_{b}^{j}F_{cd}^{kl}=i8q^{abcd}\left(
g_{bd}\theta_{c}^{i}\right)  , \label{Einst}%
\end{equation}%
\begin{equation}
\operatorname{Im}(B_{ab}\bullet B_{cd})=0,
\end{equation}
and for $S_{PT}$ we have additional equations $g_{bd}q^{abcd}=0$. Equation
(\ref{three}) simply requires the $A$ to be the Levi-Civita connection of the
metric $g_{ab}=\theta_{a}\bullet\theta_{b}$. Transforming equation
(\ref{Einst}) we get%
\begin{equation}
b(F_{ga}^{gf}-\frac{1}{2}\delta_{a}^{f}F)=-2iq^{fbcd}\left(  A_{abcd}\right)
, \label{PalEq}%
\end{equation}
where the left hand side is the Einstein tensor multiplied by $b=\det
(\theta_{a}^{i})$. In the case of $S_{cPT}$ the right hand side is zero, so
the Einstein's equations are satisfied.

Let me discuss the field equations of $S_{rPT}$. The interpretation of
equation (\ref{PalEq}) is similar to that of the various cases discussed for
the Plebanski action with the reality constraint. The right hand side is
purely imaginary because of the reality constraint. The left side is real if
1) the metric is real and the signature is Riemannian or Kleinien, 2) the
metric is imaginary and the signature is Lorentzian. So for these cases the
Einstein tensor must vanish if $b\neq0.$ So they correspond to general
relativity. For all the other combinations and also for $b=0$ the Einstein
tensor need not vanish.

\section{Conclusion}

In this article we have established a classical foundation for a concept of
reality conditions in the context of spin foam models. At the classical
continuum level it is the condition that the area metric be real. In the
context of Barrett-Crane theory\cite{JWBLC1} this takes the form of the
reality of the scalar products of the bivectors associated with the triangles
of a four simplex or three simplex. At the quantum level this idea brings
together the Barrett-Crane spin foam models of real and $SO(4,C)$ general
relativity theories in four dimensions \cite{MSK2} in a unified perspective.
In Ref:\cite{MSK2} two generalizations of real general relativity Barrett
models have been proposed. One of them puts together two Lorentzian
Barrett-Crane models to get a more general model called the mixed Lorentzian
Barrett-Crane model. Another model was defined by putting together the mixed
Lorentzian model and the Barrett-Crane models for all other signatures to get
a multi-signature model. The theory defined by the real action in equation
(\ref{Realaction}) for $SO(4,C)$ general relativity with the reality
constraint contains the general relativity for all signatures. So this theory
must be related to the multi-signature model. The precise details of this idea
need to be analyzed further. The continuum and semiclassical limits of the
various actions proposed in this article need to analyzed. Physical usefulness
need to be investigated.

\section{Acknowledgement}

I\ thank Allen Janis and George Sparling for the correspondences.

\appendix

\section{Spinorial Expansion Calculations}

Consider a tensor $R_{abcd}$ which has the symmetries of the indices of the
Riemann Curvature tensor. In this appendix I\ would like to briefly summarize
the spinorial decomposition of $R_{abcd}$. The expansion of $R_{abcd}$ in
terms of the left handed and the right handed spinorial free components
is\footnote{A suitable soldering form and a variable spinorial basis need to
be defined to map between coordinate and spinor space.}%
\begin{equation}
R_{abcd}=R_{ABCD}\epsilon_{\acute{A}\acute{B}}\epsilon_{\acute{C}\acute{D}%
}+R_{\acute{A}\acute{B}\acute{C}\acute{D}}\epsilon_{AB}\epsilon_{CD}%
+R_{AB\acute{C}\acute{D}}\epsilon_{\acute{A}\acute{B}}\epsilon_{CD}%
+R_{\acute{A}\acute{B}CD}\epsilon_{AB}\epsilon_{\acute{C}\acute{D}}.
\label{Rexpansion}%
\end{equation}
The $R_{ABCD}$ and $R_{\acute{A}\acute{B}\acute{C}\acute{D}}$ are independent
of each other and $R_{AB\acute{C}\acute{D}}=R_{\acute{C}\acute{D}AB}$ because
of the exchange symmetry. The first and last terms can be expanded into a spin
two and spin zero tensors as follows:%
\begin{align}
R_{ABCD}  &  =\frac{1}{24}R_{(ABCD)}+X(\epsilon_{AC}\epsilon_{BD}%
+\epsilon_{BC}\epsilon_{AD})~~\text{and}\label{WeylExpansion1}\\
R_{\acute{A}\acute{B}\acute{C}\acute{D}}  &  =\frac{1}{24}R_{(\acute{A}%
\acute{B}\acute{C}\acute{D})}+Y(\epsilon_{\acute{A}\acute{C}}\epsilon
_{\acute{B}\acute{D}}+\epsilon_{\acute{B}\acute{C}}\epsilon_{\acute{A}%
\acute{D}}), \label{Weyl Expansion2}%
\end{align}
where $X=\frac{1}{6}R_{AB}^{AB}$ and $Y=\frac{1}{6}R_{\acute{A}\acute{B}%
}^{\acute{A}\acute{B}}$. Let me define $C_{ABCD}=\frac{1}{24}R_{(ABCD)}$ and
$C_{\acute{A}\acute{B}\acute{C}\acute{D}}=\frac{1}{24}R_{(\acute{A}\acute
{B}\acute{C}\acute{D})}$. Let me define the two tensors%
\begin{align*}
\Pi_{ab}^{cd}  &  =\frac{1}{2}\epsilon_{ab}^{cd}\text{ and}\\
\Delta_{ab}^{cd}  &  =\frac{1}{2}\delta_{a}^{[c}\delta_{b}^{d]}.
\end{align*}
The tensor $\Delta$ is a scalar. The $\Pi_{ab}^{cd}$ is a pseudo scalar and
the dualizing operator under it's action on bivectors ($\underline{B}_{ab}%
=\Pi_{ab}^{cd}B_{cd}$ or simply $\underline{B}=\Pi B$). Let me define two
operations on $\Pi$ and $\Delta$: the product, for example $(\Pi\Delta
)_{cd}^{ab}=\Pi_{ef}^{ab}\Delta_{cd}^{ef}$ and the trace, for example
$tr(\Delta)=\Delta_{ab}^{ab}$. We can verify the following properties of $\Pi$
and $\Delta$ which are%
\begin{align*}
tr(\Pi)  &  =0,\\
tr(\Delta)  &  =6,\\
\Pi\Delta &  =\Delta\Pi=\Pi,\\
\Pi\Pi &  =\Delta~~~\text{and}\\
\Delta B  &  =B,
\end{align*}
where $B$ is an arbitrary bivector. The above properties help in the analysis
of tensor with the tensor $R_{abcd}$. The spinorial expansion of $\Pi$ and
$\Delta$ are%
\begin{align*}
\Pi_{ab}^{cd}  &  =\frac{\epsilon_{A}^{~C}\epsilon_{\grave{A}}^{~\grave{C}%
}\epsilon_{B}^{~D}\epsilon_{\grave{B}}^{~\grave{D}}-\epsilon_{A}^{~D}%
\epsilon_{\grave{A}}^{~\grave{D}}\epsilon_{B}^{~C}\epsilon_{\grave{B}%
}^{~\grave{C}}}{2}\text{ and}\\
\Delta_{ab}^{cd}  &  =\frac{\epsilon_{A}^{~D}\epsilon_{B}^{~C}\epsilon
_{\grave{A}}^{~\grave{C}}\epsilon_{\grave{B}}^{~\grave{D}}-\epsilon_{A}%
^{~D}\epsilon_{B}^{~C}\epsilon_{\grave{A}}^{~\grave{C}}\epsilon_{\grave{B}%
}^{~\grave{D}}}{2},
\end{align*}
respectively. Using equation (\ref{WeylExpansion1}) and equation
(\ref{Weyl Expansion2}) in equation (\ref{Rexpansion}) the result can be
simplified using the spinorial expansions of $\Pi$ and $\Delta$, and the
identity $\epsilon_{A[B}\epsilon_{CD]}=0$:%

\[
R_{ab}^{cd}=C_{AB}^{~CD}\epsilon_{\acute{A}\acute{B}}\epsilon^{\acute{C}%
\acute{D}}+C_{\acute{A}\acute{B}}^{~\acute{C}\acute{D}}\epsilon_{AB}%
\epsilon^{CD}+\frac{\mathcal{R}}{6}\Delta_{ab}^{cd}+\frac{\mathcal{S}}{6}%
\Pi_{ab}^{cd}+R_{AB}^{\acute{C}\acute{D}}\epsilon_{\acute{A}\acute{B}}%
\epsilon^{CD}+R_{\acute{A}\acute{B}}^{CD}\epsilon_{AB}\epsilon^{\acute
{C}\acute{D}},
\]
where $\mathcal{R}=Tr(R\Delta)=2(R_{AB}^{AB}+R_{\acute{A}\acute{B}}^{\acute
{A}\acute{B}})$ and $\mathcal{S=}Tr(R\Pi)=2(R_{AB}^{AB}-R_{\acute{A}\acute{B}%
}^{\acute{A}\acute{B}})$.

If $R_{ab}^{cd}$ is a general spatial curvature tensor then $C_{AB}^{CD}$ and
$C_{\acute{A}\acute{B}}^{\acute{C}\acute{D}}$ are the left handed and the
right handed spinorial parts of the Weyl tensor:%
\[
C_{ab}^{cd}=C_{AB}^{CD}\epsilon_{\acute{A}\acute{B}}\epsilon^{\acute{C}%
\acute{D}}+C_{\acute{A}\acute{B}}^{\acute{C}\acute{D}}\epsilon_{AB}%
\epsilon^{CD}.
\]
Each of $C_{AB}^{CD}$ and $C_{\acute{A}\acute{B}}^{\acute{C}\acute{D}}$ has
five free components. The $R_{DBB^{\prime}D^{\prime}}=-\frac{1}{2}R_{bd}$ is
the trace free Ricci tensor $R_{bd}=g^{ac}R_{abcd}-\frac{1}{4}g_{bd}R$, which
has nine free components. The $\mathcal{R}=Tr(R\Delta)$ is the scalar
curvature. The $\mathcal{S=}Tr(R\Pi)$ can be referred to as the pseudo scalar
curvature because it changes sign under the change of orientation of
space-time. It vanishes for the Riemann curvature tensor as it corresponds to
a torsion free connection. For an arbitrary curvature tensor, in terms of the
torsion the pseudo-scalar component is%
\begin{equation}
\mathcal{S}=Tr(\Pi DT), \label{Rtorsion}%
\end{equation}
where $D$ is the exterior space-time covariant derivative, $T$ is the torsion
written as a $3$-form with all of it's indices lowered and anti-symmetrized.

Under the action of dual operation $\underline{R}=\Pi R$ we have%
\[
\underline{R}_{ab}^{cd}=C_{AB}^{CD}\epsilon_{\acute{A}\acute{B}}%
\epsilon^{\acute{C}\acute{D}}-C_{\acute{A}\acute{B}}^{\acute{C}\acute{D}%
}\epsilon_{AB}\epsilon^{CD}+\frac{\mathcal{R}}{6}\Pi_{ab}^{cd}+\frac
{\mathcal{S}}{6}\Delta_{ab}^{cd}+R_{AB}^{\acute{C}\acute{D}}\epsilon
_{\acute{A}\acute{B}}\epsilon^{CD}-R_{\acute{A}\acute{B}}^{CD}\epsilon
_{AB}\epsilon^{\acute{C}\acute{D}}.
\]
Notice that the $R_{AB}^{\acute{C}\acute{D}}$ and $R_{\acute{A}\acute{B}}%
^{CD}$ terms have different signs, $\mathcal{R}$ and $\mathcal{S}$ exchanged
positions. These properties are crucial for interpreting the field equation
(\ref{Feq}) of the Plebanski formulation of general relativity.

\end{document}